\documentclass[8pt]{revtex4}
\pdfoutput=1
\usepackage{amssymb}
\usepackage{latexsym}
\usepackage{epsfig}
\begin{document}                             

\title{ Cosmology from quantum potential in brane-anti-brane system}

\author{ Alireza Sepehri \footnote{alireza.sepehri@uk.ac.ir} }
\address{ Faculty of Physics,
Shahid Bahonar University, P.O. Box 76175, Kerman, Iran.}

\begin{abstract}
Recently, some authors removed the big-bang singularity and
predicted an infinite age of our universe. In this paper, we show
that the same result can be obtained in string theory and
M-theory; however, the shape of universe changes in different
epochs. In our mechanism, first,  N fundamental string decay to N
D0-anti-D0-brane. Then, D0-branes join to each other, grow and and
form a six-dimensional brane-antibrane system. This system is
unstable, broken and  present form of  four dimensional universes
, one anti-universe in additional to one wormhole are produced.
Thus, there isn't any big-bang in cosmology and universe is a
fundamental string at the beginning. Also, total age of universe
contains two parts, one in related to initial age and second which
is corresponded to present age of universe
($t_{tot}=t_{initial}+t_{present}$). On the other hand, initial
age of universe includes two parts, the age of fundamental string
and time of transition
($t_{initial}=t_{transition}+t_{f-string}$). We observe that only
in the case of ($t_{f-string}\rightarrow \infty$), the scale
factor of universe is zero and as a result, total age of universe
is infinity.

\end{abstract}

 \maketitle
\section{Introduction}
 Recently, some authors obtained the second order Friedmann equations from the  quantum corrected Raychaudhuri equation (QRE),
  and argued that these equations include two quantum correction terms,
   the first of which can be interpreted as cosmological constant,
    while the second as a radiation term in the early universe,
    which gets rid of the big-bang singularity and predicts an infinite age of our universe
    \cite{q1}.  In parallel, some models have been proposed to remove
    big-bang singularity in BIonic system \cite{q6,q7,q8,q9}. In this system, at the beginning, there
    are k black fundamental strings that transited to the BIon
configuration at an special energy. At this energy, two brane and
antibrane universes are produced, interact with each other through
a wormhole and inflate. This wormhole gives it's energy to the
brane universes and causes to inflation. Shortly, the wormhole
disappears, the inflation ends and a deceleration epoch begins.
When the brane and antibrane universes become close together, a
tachyon is produced, grows and leads the creation of a new
wormhole. At this stage, this tachyonic wormhole connects the
brane and antibrane universes again  and leads to the late-time
acceleration era of the universe \cite{q6,q8}.

Now, the main question arises as to what is the relation between
the model which was obtained from QRE and the one which was used
BIon?  We will answer this question in this paper. We also modify
previous discussions in BIon by introducing D0-branes and clarify
transitions in this system.  In our model,  at the beginning,
there are N fundamental strings that excited and make a transition
 to N D0-anti-D0-branes. Then, by joining and growing these objects, a pair of brane-antibrane is formed.
 These branes interact with each other, emit D0-branes and changes to lower dimensional branes like D3-branes.
  The present universe is located on one of these D3-branes and radiated D0-branes form a
   wormhole that connects another universe with our one. This system which includes two universe-branes with one wormhole connected them,
    is named BIon.  Ignoring evolution of universe, before formation of
    BIon, the scale factor is zero at the beginning. However,
    regarding the transition of universe from fundamental string to
    BIon, the scale factor is zero only in the case of $t_{universe}\rightarrow \infty$ and the age of universe is
    infinite.

Maybe the question arises as to why does  one choose this rather
complicated model of our Universe (string to branes, to Universe,
anti-universe and wormhole)? To reply this question, we can say
that our model provides good reasons  for  the small size of
Universe at the beginning and then enlarging  to the current size.
According to resent observations  \cite{R1,R2,R3},  at  very early
epoch, universe was very small and experienced a phase of
inflation.  In our model,  the   initial fundamental string is
very tiny  and then decays  to little D0-branes. These branes
construct  two small Universe-anti-Universe-branes and one
wormhole between them. This wormhole gives it's energy to
D3-branes and dissolves into them and causes to their expansion.
Thus, this model helps us to consider evolution of Universe in
string theory.

Also, one may ask that how does one know that the initial state of
our Universe is a fundamental string?  To answer this question, we
emphasis that the origin of  BIon  is the fundamental string
\cite{q6,q8}. Previously,  it has been shown that all evolution of
Universe can be described in a Bionic system \cite{q6,q7,q8,q9}.
Thus, if our Universe  be a part of BIon , the initial state of it
should be a fundamental string . Also, the small size of initial
Universe is comparable with the size of fundamental string. Thus,
we can remove the big bang and show that all evolution of the
Universe begins from the fundamental string in a BIon.

 The outline of the
paper is as  follows.  In section \ref{o1}, we will remove the Big
bang singularity and show that the origin of universe is a
fundamental string in a BIonic system. In section \ref{o2}, we
will extend our calculations to M-theory and show that the age of
universe is infinite. The last section is devoted to summary and
conclusion.

\section{ Removing the big-bang singularity in BIon}\label{o1}
In this section, we will propose a new model which allows to
replace big-bang singularity by a fundamental string. In this
model the present form of universe is created via process
(fundamental string $\rightarrow$ D0 + anti-D0 $\rightarrow$ D5 +
anti-D5 $\rightarrow$ D3 + anti-D3  $\rightarrow$ universe +
anti-universe + wormhole). There isn't any birth time for the
fundamental string and thus, the age of universe could be
infinite.

First, let us to introduce the mechanism of \cite{q1} in short
terms. In this method, we calculate an explicit form for
$\dot{H}=F(H)$, where F(H) is a function of Hubble parameter (H)
that is obtained from QRE. Using this function, we can find the
age of universe:

\begin{eqnarray}
&& \dot{H}= F(H)\rightarrow T= \frac{1}{F^{n}(H_{initial})}\int dH
\frac{1}{(H-H_{initial})^{n}}\rightarrow \infty\label{a4}
\end{eqnarray}

where $H_{initial}$ is the hubble parameter before the present
epoch of universe. As can be seen from this equation, the age of
universe is infinity. To achieve this result in string theory, at
first stage, we should calculate the energy density and momentums
of a universe + anti-universe + wormhole system. With respect to
the FRW metric,

\begin{eqnarray}
&& ds^{2}_{FRW}  = -dt^{2} + a(t)^{2}(dx^{2} +
dy^{2} + dz^{2}), \label{a4}
\end{eqnarray}

we can calculate the energy density and pressure in one flat
universe\cite{q10}:

\begin{eqnarray}
&& \rho_{uni} = 3 H^{2},
 \nonumber \\ && p_{uni} = H^{2} + 2\frac{\ddot{a}}{a} \label{a5}
\end{eqnarray}

Where $H=\frac{\dot{a}}{a}$ is the Hubble parameter and a is the scale factor.

To consider the evolution of universe in BIonic system, we
introduce two four dimensional universes that interact with each
other via a wormhole and form a binary system. In this model, our
 universe is located on one D-brane and connected by another
universe on the anti-brane by a wormhole. The metric of wormhole
is given by \cite{q11}:
\begin{eqnarray}
&& ds_{wormhole}^{2} =
B(r)(-dt^{2}+dr^{2})+C(r)d\phi^{2}+D(r)du^{2}\label{a6}
\end{eqnarray}
in which B (r), C (r) and D(r) are as a function of r only and r=V(t)
is the location of the throat. The standard energy momentum on the
shell of wormhole is:
\begin{eqnarray}
&& \rho_{wormhole} = - (\frac{D'}{D}+\frac{C'}{C})\sqrt{\Delta}\nonumber\\
&& p_{u,wormhole}=\frac{1}{\sqrt{\Delta}}[2\ddot{V}+2\frac{B'}{B}\dot{V}^{2}+\frac{B'}{B^{2}}+\frac{C'}{C}\Delta],\nonumber\\
&&
p_{\phi,wormhole}=\frac{1}{\sqrt{\Delta}}[2\ddot{V}+2\frac{B'}{B}\dot{V}^{2}+\frac{B'}{B^{2}}+\frac{D'}{D}\Delta]
\label{a7}
\end{eqnarray}
where $\Delta=\frac{1}{B}+\dot{V}^{2}$.

The general metric of this system is:

\begin{eqnarray}
&&ds_{uni-wormhole}^{2} =ds_{uni1}^{2}+ds_{uni2}^{2}+ds_{wormhole}^{2} \label{a8}
\end{eqnarray}

According to conservation law, the energy and momentums of this system should be equal to energy and momentums of brane-anti-brane system.
We should write:

\begin{eqnarray}
&&\rho_{brane-anti-brane}= \rho_{uni1}+ \rho_{uni2}+ \rho_{wormhole}
\nonumber\\&&p_{brane-anti-brane}= p_{uni1}+ p_{uni2}+ p_{wormhole}
\label{a9}
\end{eqnarray}

Now, we need to compute the contribution of the BIonic system with
the four- dimensional universe energy momentum tensor. To obtain
this tensor, we use of following action for
\cite{q12,q13,q14,q15,qq15,q16}:

\begin{eqnarray}
S = - T_{Dp}\int  d^{p+1}\sigma ~ STr \Bigg(-det(P_{ab}[E_{mn}  E_{mi}(Q^{-1}+\delta)^{ij}E_{jn}]+ \lambda F_{ab})det(Q^{i}_{j})\Bigg)^{1/2}~~
\label{a10}
\end{eqnarray}

where

\begin{eqnarray}
   E_{mn} = G_{mn} + B_{mn}, \qquad  Q^{i}_{j} = \delta^{i}_{j} + i\lambda[X^{j},X^{k}]E_{kj} \label{a11}
\end{eqnarray}

 $\lambda=2\pi l_{s}^{2}$, $G_{ab}=\eta_{ab}+\partial_{a}X^{i}\partial_{b}X^{i}$ and $X^{i}$ are scalar fields of mass dimension. Here $a,b=0,1,...,p$
are the world-volume indices of the Dp-branes, $i,j,k = p+1,...,9$
are indices of the transverse space, and m,n are the
ten-dimensional spacetime indices. Also, $T_{Dp}=\frac{1}{g_{s}(2\pi)^{p}l_{s}^{p+1}}$
is the tension of Dp-brane, $l_{s}$ is the string length and $g_{s}$ is the string coupling.
 We can approximately obtain a simple form for the action of Dp  brane \cite{q12,q13,q14,q15,q16}:

\begin{eqnarray}
&& S_{Dp} = -T_{Dp} \int d^{p+1}\sigma Tr
(\Sigma_{a,b=0}^{p}
\Sigma_{i,j=p+1}^{9}
\{\partial_{a}X^{i}\partial_{b}X^{i}-\frac{1}{2 \lambda^{2}}[X^{i},X^{j}]^{2}+\frac{\lambda^{2}}{4}
(F_{ab})^{2}
\})
\label{a12}
\end{eqnarray}

Using  following rules \cite{q12,q13,q14,q15,qq15,q16}:

\begin{eqnarray}
&& \Sigma_{a=0}^{p}\Sigma_{m=0}^{9}\rightarrow \frac{1}{(2\pi l_{s})^{p}}\int d^{p+1}\sigma \Sigma_{m=p+1}^{9}\Sigma_{a=0}^{p} \qquad \lambda = 2\pi l_{s}^{2} \nonumber \\
&&[X^{a},X^{i}]=i
\lambda \partial_{a}X^{i}\qquad  [X^{a},X^{b}]= i \lambda^{2} F^{ab}\nonumber \\
&& i,j=p+1,..,9\qquad a,b=0,1,...p\qquad m,n=0,1,..,9
\label{a13}
\end{eqnarray}

in action (\ref{a12}) and after  some mathematical calculations, we get:

\begin{eqnarray}
&& S_{Dp} = -T_{Dp} \int d^{p+1}\sigma Tr
(\Sigma_{a,b=0}^{p}
\Sigma_{i,j=p+1}^{9}
\{\partial_{a}X^{i}\partial_{b}X^{i}-\frac{1}{2 \lambda^{2}}[X^{i},X^{j}]^{2}+\frac{\lambda^{2}}{4}
(F_{ab})^{2}
\})= \nonumber \\ &&-\Sigma_{a=0}^{p}T_{D0}
\int dt Tr(
\Sigma_{m=0}^{9}
[X^{m},X^{n}]^{2}) = \Sigma_{a=0}^{p}S_{D0}
\label{a14}
\end{eqnarray}

where we have used of the action of
D0-brane\cite{q12,q13,q14,q15,qq15,q16}:

\begin{eqnarray}
&& S_{D0} =
-T_{D0}
\int dt Tr(
\Sigma_{m=0}^{9}
[X^{m},X^{n}]^{2})
\label{a15}
\end{eqnarray}

Here $T_{D0}$ is the brane tension and $X^{m}$ are transverse
scalars. The origin of this D0-brane is a fundamental string.  We
can write:

\begin{eqnarray}
&& T_{F-string}=\frac{1}{2\pi l_{s}} \qquad T_{D1}=\frac{1}{2\pi g_{s}l_{s}}\Rightarrow\nonumber \\ &&g_{s}=1 \rightarrow T_{F-string}=T_{D1} \qquad \text{and} \qquad  S_{D0}=S_{anti-D0} \Rightarrow \nonumber \\ && S_{F-string} = -T_{F-string} \int d^{p+1}\sigma Tr
(\Sigma_{a,b=0}^{1}
\Sigma_{i,j=2}^{9}
\{\partial_{a}X^{i}\partial_{b}X^{i}-\frac{1}{2 \lambda^{2}}[X^{i},X^{j}]^{2}+\frac{\lambda^{2}}{4}
(F_{ab})^{2}
\})= \nonumber \\ &&-\Sigma_{a=0}^{1}T_{D0}
\int dt Tr(
\Sigma_{m=0}^{9}
[X^{m},X^{n}]^{2}) = \Sigma_{a=0}^{1}S_{D0}= S_{D0}+S_{anti-D0}
\label{aa14}
\end{eqnarray}

This equation indicates that at strong coupling, fundamental string is broken and two D0-anti-D0-branes are produced.  Using the action (\ref{a14}), we can propose a new mechanism for interaction of two brane-anri-brane. We can write:

\begin{eqnarray}
&& S_{Dp} + S_{anti-Dp}= \Sigma_{a=0}^{p}S_{D0} +
\Sigma_{a=0}^{p}S_{D0}= \nonumber \\&&\Sigma_{a=0}^{p-2}S_{D0} +
\Sigma_{a=0}^{p-2}S_{D0}+ \Sigma_{a=0}^{2}S_{D0}+S_{D0}=\nonumber
\\&&S_{D(p-2)} + S_{anti-D(p-2)}+S_{D2} +S_{D0}\label{a16}
\end{eqnarray}

This equation shows that each of Dp-branes in a brane-antibrane system emit two D0-branes and transits to a lower dimensional D(p-2) brane and one D2-brane is produced. For example, a system of D5-anti-D5-brane is broken and a system of D3-anti-D3-D2-brane is formed. To construct a D3-anti-D3-D2 model, we consider two
D3-anti-D3-brane pairs  which
are placed at points $u_{1} = l_{2}/2$ and $u_{2} = -l_{2}/2$ respectively
so that the separation between the brane and antibrane is $l_{2}$. Here, $l_{2}$ is the length of D2-brane in transverse direction respect to D3-brane. Following rules in (\ref{a13}), we can obtain the solutions for  gauge fields in D3 and D2-branes \cite{q12,q13,q14}:

\begin{eqnarray}
&&  \lambda F_{01} \text{ in D2-brane } \rightarrow \partial_{t} X^{1} \text{ in D3-brane} \qquad \nonumber \\&& F_{ab} \text{ in D3-brane } \rightarrow [X^{a},X^{b}]\text{ in D2-brane }\Rightarrow \nonumber \\&&
X^{i} \sim \frac{l_{3}}{2},\quad A^{i} \sim \frac{l_{2}}{2} \qquad \text{ in D2-brane }
\nonumber \\&& X^{i} \sim \frac{l_{2}}{2},\quad
A^{i} \sim  \frac{l_{3}}{2} \qquad \text{ in D3-brane }
\label{a17}
\end{eqnarray}

where $l_{2}$ and $l_{3}$ are coordinates of D2 and D3-branes respectively. By passing time, $l_{2}$ is decreased and  reduced to zero at the end of acceleration and   $l_{3}$ is increased. Using these rules and action (\ref{a11}), we can write the following equation for this  D3- brane-anti-D3-brane+ D2-brane system:

\begin{eqnarray}
&& S_{D3-D2-anti-D3}=S_{D3}+S_{anti-D3}+S_{D2} +S_{D0}\simeq
\nonumber \\ && -T_{D3} \int d^{4}\sigma
(l_{2})^{4}(\sqrt{1+\frac{(l'_{2})^{2}}{4}+\frac{(l'_{3})^{2}}{4}}+\sqrt{1+\frac{(l'_{2})^{2}}{4}+\frac{(l'_{3})^{2}}{4}})\nonumber
\\ && -T_{D2} \int d^{3}\sigma
(l_{3})^{4}(\sqrt{1+\frac{(l'_{2})^{2}}{4}+\frac{(l'_{3})^{2}}{4}})
 \nonumber
\\ && -T_{D0} \int d\sigma
((l_{3})^{4}+(l_{2})^{4})(\sqrt{1+\frac{(l'_{2})^{2}}{4}+\frac{(l'_{3})^{2}}{4}})\simeq\nonumber
\\ && -\int dt \frac{(l_{3})^{3}(l_{2})^{2}}{\pi g_{s}l_{s}^{3}}
(\frac{2(l_{2})^{2}}{\pi
l_{s}}+l_{3}+l_{3}l_{2}^{-2}+l_{2}^{2}l_{3}^{-3})(\sqrt{1+\frac{(l'_{2})^{2}}{4}+\frac{(l'_{3})^{2}}{4}})
\simeq\nonumber
\\ && -\int dt F_{l_{2},l_{3}}(\sqrt{D_{l_{2},l_{3}}}) \label{a18}
\end{eqnarray}

where
\begin{eqnarray}
   F_{l_{2},l_{3}}= \frac{(l_{3})^{3}(l_{2})^{2}}{\pi g_{s}l_{s}^{3}}
(\frac{2(l_{2})^{2}}{\pi
l_{s}}+l_{3}+l_{3}l_{2}^{-2}+l_{2}^{2}l_{3}^{-3}), \qquad
D_{l_{2},l_{3}}=1+\frac{(l'_{2})^{2}}{4}+\frac{(l'_{3})^{2}}{4}\label{a19}
\end{eqnarray}

and prime($'$) denotes derivative respect to time (t). To obtain the explicit form
of $l_{2}$ and $l_{3}$ in terms of time , we are using the equation of motion extracted from
action (\ref{a18}):
\begin{eqnarray}
(\frac{l_{2}'(t)}{\sqrt{1+\frac{(l'_{2})^{2}}{4}+\frac{(l'_{3})^{2}}{4}}}\acute{)}=\frac{1}{\sqrt{1+\frac{(l'_{2})^{2}}{4}+\frac{(l'_{3})^{2}}{4}}}
[\frac{F'_{l_{2},l_{3}}}{F_{l_{2},l_{3}}}(D_{l_{2},l_{3}}-(l_{2}'(t))^{2})]
\label{a20}
\end{eqnarray}

\begin{eqnarray}
(\frac{l_{3}'(t)}{\sqrt{1+\frac{(l'_{2})^{2}}{4}+\frac{(l'_{3})^{2}}{4}}}\acute{)}=\frac{1}{\sqrt{1+\frac{(l'_{2})^{2}}{4}+\frac{(l'_{3})^{2}}{4}}}
[\frac{F'_{l_{2},l_{3}}}{F_{l_{2},l_{3}}}(D_{l_{2},l_{3}}-(l_{3}'(t))^{2})]
\label{a21}
\end{eqnarray}

Solving above equations, we obtain:
\begin{eqnarray}
&& l_{3}\simeq t^{2}(1+e^{\frac{(t-t_{end})}{\pi l_{s}}})^{1/3}\nonumber \\ &&l_{2}\simeq t^{3}(1-e^{\frac{(t-t_{end})}{\pi l_{s}}})^{1/2}
\label{a22}
\end{eqnarray}

where  $t_{end}$ is the age of universe-brane at the end of epoch. An interesting result that comes out of these equation is that the length of D2-brane in transverse dimension increases with time, turns over a maximum and then decreases  and shrinks to zero at $t_{end}$, while; D3-brane is expanded very fast.  The energy momentum tensor for this system is \cite{q6,q8,q9}:

 \begin{eqnarray}
&& T^{00}=F_{l_{2},l_{3}}\sqrt{D_{l_{2},l_{3}}},  \nonumber \\&&
T^{44}=-F_{l_{2},l_{3}}\frac{1}{3\sqrt{D_{l_{2},l_{3}}}}
((l_{3})^{2}(l_{2})^{2}+\frac{(l'_{2})^{2}}{4}) \nonumber \\&& T^{ii} =
-F_{l_{2},l_{3}}\frac{1+(l_{3})^{2}}{\sqrt{D_{l_{2},l_{3}}}},\,i=1,2,3 \label{a23}
\end{eqnarray}

 This equation indicates that with the energy-momentum tensors increases with time. This is
due to the fact that with approaching two D3-branes each other,  D2-brane dissolves in two universe branes and causes to expansion and and increase in values of density and momentum of universes. This higher-dimensional stress-energy tensor
is related to a perfect fluid and of the form
 \begin{equation}
T_i^j = {\mathop{\rm diag}\nolimits} \left[ { - p, - p, - p, -
\bar{p}, - p, - p, - p, \rho } \right], \label{a24}
\end{equation}
 where $\bar{p}$ is the pressure in the extra space-like
 (u) dimension.  Thus, this relation and also conversation low in equation  (\ref{a9}) help us to write:

\begin{eqnarray}
&&\rho_{brane-anti-brane}= \rho_{uni1}+ \rho_{uni2}+ \rho_{wormhole}
\Rightarrow \nonumber
\\&&6H_{present}^{2} - (\frac{D'}{D}+\frac{C'}{C})\sqrt{\Delta}=\nonumber
\\&& F_{l_{2},l_{3}}\sqrt{D_{l_{2},l_{3}}} \nonumber
\\&&\nonumber
\\&&p_{brane-anti-brane,i}= p_{uni1,i}+ p_{uni2,i}+ p_{wormhole,i}
\Rightarrow \nonumber
\\&&
2H_{present}^{2} +
4\frac{\ddot{a}_{present}}{a_{present}}+\frac{1}{\sqrt{\Delta}}[2\ddot{V}+2\frac{B'}{B}\dot{V}^{2}+\frac{B'}{B^{2}}+\frac{D'}{D}\Delta]
=\nonumber
\\&&
-F_{l_{2},l_{3}}\frac{1+(l_{3})^{2}}{3\sqrt{D_{l_{2},l_{3}}}}
\nonumber
\\&& \nonumber
\\&&p_{brane-anti-brane,u}= p_{uni1,u}+ p_{uni2,u}+ p_{wormhole,u}
\Rightarrow \nonumber
\\&&\frac{1}{\sqrt{\Delta}}[2\ddot{V}+2\frac{B'}{B}\dot{V}^{2}+\frac{B'}{B^{2}}+\frac{C'}{C}\Delta]=\nonumber
\\&&-F_{l_{2},l_{3}}\frac{1}{\sqrt{D_{l_{2},l_{3}}}}
((l_{3})^{2}(l_{2})^{2}+\frac{(l'_{2})^{2}}{4}) \label{a25}
\end{eqnarray}

where the index present refers to present stage of universe.
Assuming $D=C$ and $V=l_{2}$ , we can derive the solutions of
these equations:

\begin{eqnarray}
&& a_{present}(t)=  e^{-\int dt \ln(X^{-1})},\qquad C(t)=D=
e^{-\int dt \ln(Y^{-1})}, \qquad D=e^{-\int dt \ln(Z^{-1})}
\nonumber
\\&& \nonumber
\\&& X=[t^{10}(1+e^{\frac{(t-t_{end})}{\pi
l_{s}}})(1-e^{\frac{(t-t_{end})}{\pi
l_{s}}})][t^{3/2}(1-e^{\frac{(t-t_{end})}{\pi
l_{s}}})+t^{1/2}(1+e^{\frac{(t-t_{end})}{\pi
l_{s}}})]\times\nonumber\\&&  [1+(2t(1+e^{\frac{(t-t_{end})}{\pi
l_{s}}})^{1/3}+t^{2}(t-t_{end})e^{\frac{(t-t_{end})}{\pi
l_{s}}}(1+e^{\frac{(t-t_{end})}{\pi l_{s}}})^{-2/3})^{2}+\nonumber
\\&& (3t^{2}(1-e^{\frac{(t-t_{end})}{\pi l_{s}}})^{1/2}+t^{3}(t-t_{end})e^{\frac{(t-t_{end})}{\pi l_{s}}}(1-e^{\frac{(t-t_{end})}{\pi l_{s}}})^{-1/2})^{2}]^{1/4}\nonumber
\\&& \nonumber
\\&&Y= ([2t(t_{end}-t)e^{\frac{(t-t_{end})}{\pi l_{s}}}(1-e^{\frac{(t-t_{end})}{\pi l_{s}}})^{-2/3}
 +2t(1-e^{\frac{(t-t_{end})}{\pi l_{s}}})^{1/3}]+\nonumber\\&&[t^{10}(1+e^{\frac{(t-t_{end})}{\pi
l_{s}}})(1-e^{\frac{(t-t_{end})}{\pi
l_{s}}})][t^{3/2}(1-e^{\frac{(t-t_{end})}{\pi
l_{s}}})+t^{1/2}(1+e^{\frac{(t-t_{end})}{\pi
l_{s}}})])\times\nonumber\\&& [t^{9}(1-e^{\frac{(t-t_{end})}{\pi
l_{s}}})(1+e^{\frac{(t-t_{end})}{\pi
l_{s}}})^{4/3}(3t(1-e^{\frac{(t-t_{end})}{\pi
l_{s}}})^{1/2}+t^{3}(t-t_{end})e^{\frac{(t-t_{end})}{\pi
l_{s}}}(1-e^{\frac{(t-t_{end})}{\pi l_{s}}})^{-1/2})^{2}]\times
\nonumber
\\&&[1+(2t(1+e^{\frac{(t-t_{end})}{\pi l_{s}}})^{1/3}+t^{2}(t-t_{end})e^{\frac{(t-t_{end})}{\pi l_{s}}}(1+e^{\frac{(t-t_{end})}{\pi l_{s}}})^{-2/3})^{2}+\nonumber
\\&& (3t^{2}(1-e^{\frac{(t-t_{end})}{\pi l_{s}}})^{1/2}+t^{3}(t-t_{end})e^{\frac{(t-t_{end})}{\pi l_{s}}}(1-e^{\frac{(t-t_{end})}{\pi l_{s}}})^{-1/2})^{2}]^{-1/2} \nonumber
\\&&\nonumber
\\&& Z= [1+t^{4}(1+e^{\frac{(t-t_{end})}{\pi l_{s}}})^{2/3}]\times \nonumber
\\&&[t^{10}(1+e^{\frac{(t-t_{end})}{\pi
l_{s}}})(1-e^{\frac{(t-t_{end})}{\pi
l_{s}}})][t^{3/2}(1-e^{\frac{(t-t_{end})}{\pi
l_{s}}})+t^{1/2}(1+e^{\frac{(t-t_{end})}{\pi
l_{s}}})]\times\nonumber\\&&[1+(2t(1+e^{\frac{(t-t_{end})}{\pi
l_{s}}})^{1/3}+t^{2}(t-t_{end})e^{\frac{(t-t_{end})}{\pi
l_{s}}}(1+e^{\frac{(t-t_{end})}{\pi l_{s}}})^{-2/3})^{2}+\nonumber
\\&& (3t^{2}(1-e^{\frac{(t-t_{end})}{\pi l_{s}}})^{1/2}+t^{3}(t-t_{end})e^{\frac{(t-t_{end})}{\pi l_{s}}}(1-e^{\frac{(t-t_{end})}{\pi l_{s}}})^{-1/2})^{2}]^{-1/2} \label{a26}
\end{eqnarray}

This equation indicates that by passing time, the parameters of
wormhole and four dimensional universe are increased, turn over a
maximum and then reduced to lower values and tended to zero at
$t=t_{end}$. This means that universe is born at one beginning
time (t=0), expands in BIonic system, then contracts and vanishes
at $t=t_{end}$.

Now, we add contributions of initial stages which fundamental
string transits to BIon. To this end, we define a new scale factor
$a_{tot}=a_{initial}+a_{present}$ which $a_{initial}$ can be
obtained from following equation:

\begin{eqnarray}
&&\rho_{\text{fundamental string}}=\rho_{3D0}+ \rho_{3-anti-D0}=
\rho_{uni1}+ \rho_{uni2}\Rightarrow \nonumber
\\&&6H_{initial}^{2} = 2T_{D0}
\int dt Tr( \Sigma_{m=0}^{9} [X^{m},X^{n}]^{2})\simeq
\frac{t^{5}}{5}(1+e^{\frac{(t-t_{end})}{\pi
l_{s}}})+t^{4}(t-t_{end})^{2}\rightarrow \nonumber
\\&& a_{initial}(t)=Aexp(-\int_{t_{f-string}}^{t} dt {\frac{t^{5}}{5}(1+e^{\frac{(t-t_{end})}{\pi
l_{s}}})+t^{4}(t-t_{end})^{2}})
 \label{aa25}
\end{eqnarray}

where, $t_{f-string}$ is the age of the fundamental string and we
have used of this fact that $X^{2}=l_{2}$ and $X^{3}=l_{3}$. When
($t_{f-string}\rightarrow \infty$), the initial scale factor is
zero. Thus, total age of universe contains two parts, one in
related to initial age and second which is corresponded to present
age of universe ($t_{tot}=t_{initial}+t_{present}$). Also, initial
age of universe includes the age of fundamental string and time of
transition ($t_{initial}=t_{transition}+t_{f-string}$). As a
result of ($t_{f-string}\rightarrow \infty$), total age of
universe is infinity.

\section{Removing the big bang in M-Theory }\label{o2}
In this section, we will replace D0 by M0 in M-theory and show
that this object has the main role in evolution of universe. In
this model, first fundamental string transits to M0-branes and
anti-M0-branes and then these objects join to each and form a
M5-anti-M5-system. The branes in this system, emit M0-branes,
broken and transits to lower dimensional M3 and anti-M3-brane.
Coincidence with the birth of this new system, M0-branes glued to
each other and construct a wormhole. This wormhole connects two
universes that are located on each M3-brane. As a result the
initial shape of M3-branes and universes is a fundamental string.
This string has no the birth and for this reason, the real age of
universe is infinity.

 Let us to begin with
the Born-Infeld action for M0-brane by replacing two dimensional
Nambu-Poisson bracket \cite{q17,q19,q21,q22} for Dp-branes by
three one in action and using the Li-3-algebra:

\begin{eqnarray}
S_{M0} =
T_{M0}\int dt Tr(
\Sigma_{M,N,L=0}^{10}
\langle[X^{M},X^{N},X^{L}],[X^{M},X^{N},X^{L}]\rangle)
\label{a29}
\end{eqnarray}

where $X^{M}=X^{M}_{\alpha}T^{\alpha}$ and

\begin{eqnarray}
 &&[T^{\alpha}, T^{\beta}, T^{\gamma}]= f^{\alpha \beta \gamma}_{\eta}T^{\eta} \nonumber \\&&\langle T^{\alpha}, T^{\beta} \rangle = h^{\alpha\beta} \nonumber \\&& [X^{M},X^{N},X^{L}]=[X^{M}_{\alpha}T^{\alpha},X^{N}_{\beta}T^{\beta},X^{L}_{\gamma}T^{\gamma}]\nonumber \\&&\langle X^{M},X^{M}\rangle = X^{M}_{\alpha}X^{M}_{\beta}\langle T^{\alpha}, T^{\beta} \rangle
\label{aa29}
\end{eqnarray}

where  $X^{M}$(i=1,3,...10) are transverse scalars to M0-brane. By compactifying M-theory on a circle of radius R,
this action will be made a transition to ten dimensional action for D0-brane. To show this, we use of the method in \cite{q22}
and define $<X^{10}>=\frac{R}{l_{p}^{3/2}}$ where $l_{p}$ is the Planck length. We have:

\begin{eqnarray}
&& S_{M0} = -
T_{M0}\int dt Tr(
\Sigma_{M,N,L=0}^{10}
\langle[X^{M},X^{N},X^{L}],[X^{M},X^{N},X^{L}]\rangle)) = \nonumber \\
&& - T_{M0}\int dt Tr(\Sigma_{M,N,L,E,F,G=0}^{10}\varepsilon_{MNLD}\varepsilon_{EFG}^{D}X^{M}X^{N}X^{L}X^{E}X^{F}X^{G} = \nonumber \\
&& - 6T_{M0}\int dt Tr(\Sigma_{M,N,E,F=0}^{9}\varepsilon_{MN10D}\varepsilon_{EF10}^{D}X^{M}X^{N}X^{10}X^{E}X^{F}X^{10} - \nonumber \\
&& 6T_{M0} \int dt \Sigma_{M,N,L,E,F,G=0,\neq 10}^{9}\varepsilon_{MNLD}\varepsilon_{EFG}^{D}X^{M}X^{N}X^{L}X^{E}X^{F}X^{G} = \nonumber \\
&& - 6T_{M0}(\frac{R^{2}}{l_{p}^{3}})\int dt Tr(\Sigma_{M,N,E,F=0}^{9}\varepsilon_{MN10D}\varepsilon_{EF10}^{D}X^{M}X^{N}X^{E}X^{F} - \nonumber \\
&& 6T_{M0}\int dt \Sigma_{M,N,L,E,F,G=0,\neq10}^{9}\varepsilon_{MNLD}\varepsilon_{EFG}^{D}X^{M}X^{N}X^{L}X^{E}X^{F}X^{G} = \nonumber \\
&& - 6T_{M0}(\frac{R^{2}}{l_{p}^{3}})\int dt Tr(\Sigma_{M,N=0}^{9}[X^{M},X^{N}]^{2}) - \nonumber \\
&& 6T_{M0} \int dt \Sigma_{M,N,L,E,F,G=0,\neq 10}^{9}\varepsilon_{MNLD}\varepsilon_{EFG}^{D}X^{M}X^{N}X^{L}X^{E}X^{F}X^{G}= \nonumber \\
&& S_{D0} -  6T_{M0} \int dt \Sigma_{M,N,L,E,F,G=0,\neq10}^{9}\varepsilon_{MNLD}\varepsilon_{EFG}^{D}X^{M}X^{N}X^{L}X^{E}X^{F}X^{G}\nonumber \\
&& S_{D0} + V_{Extra,1}
\label{a30}
\end{eqnarray}

where  $T_{M0/D0}$  is tension of brane and $ V_{Extra,1}=
-6T_{M0}\int dt
\Sigma_{M,N,L,E,F,G=0}^{9}\varepsilon_{MNLD}\varepsilon_{EFG}^{D}X^{M}X^{N}X^{L}X^{E}X^{F}X^{G}$.
We define
$T_{D0}=6T_{M0}(\frac{R^{2}}{l_{p}^{3}})=\frac{1}{g_{s}l_{s}}$
where $g_{s}$ and $l_{s}$ are the string coupling and string
length respectively. Thus, the actions in string theory and
M-theory are completely related and all results in string theory
can be generalized to M-theory.

Similar to Dp-branes, different Mp-branes can be built from
M0-brane by using the following rules  \cite{q17,q19,q21,q22}:

\begin{eqnarray}
&&\langle[X^{a},X^{b},X^{i}],[X^{a},X^{b},X^{i}]\rangle=
\frac{1}{2}\varepsilon^{abc}\varepsilon^{abd}(\partial_{a}X^{i}_{\alpha})(\partial_{a}X^{i}_{\beta})\langle(T^{\alpha},T^{\beta}\rangle =
 \frac{1}{2}\langle \partial_{a}X^{i},\partial_{a}X^{i}\rangle \nonumber \\
&&\nonumber \\
&&\langle[X^{a},X^{b},X^{c}],[X^{a},X^{b},X^{c}]\rangle=
(F^{abc}_{\alpha\beta\gamma})(F^{abc}_{\alpha\beta\eta})\langle[T^{\alpha},T^{\beta},T^{\gamma}],[T^{\alpha},T^{\beta},T^{\eta}]\rangle)=\nonumber \\
&& (F^{abc}_{\alpha\beta\gamma})(F^{abc}_{\alpha\beta\eta})f^{\alpha \beta \gamma}_{\sigma}h^{\sigma \kappa}f^{\alpha \beta \eta}_{\kappa} \langle T^{\gamma},T^{\eta}\rangle=
(F^{abc}_{\alpha\beta\gamma})(F^{abc}_{\alpha\beta\eta})\delta^{\kappa \sigma} \langle T^{\gamma},T^{\eta}\rangle=
\langle F^{abc},F^{abc}\rangle\nonumber \\
&&\nonumber \\
&&\Sigma_{m}\rightarrow \frac{1}{(2\pi)^{p}}\int d^{p+1}\sigma \Sigma_{m-p-1}
i,j=p+1,..,10\quad a,b=0,1,...p\quad m,n=0,..,10~~
\label{a31}
\end{eqnarray}

where

\begin{eqnarray}
&&F_{abc}=\partial_{a} A_{bc}-\partial_{b} A_{ca}+\partial_{c} A_{ab}\label{a32}
\end{eqnarray}

and $A_{ab}$ is 2-form gauge field. Replacing  commutation relations
by derivatives and fields of equations (\ref{a31})
in action (\ref{a29}), we can obtain the relevant action for Mp-brane

\begin{eqnarray}
&& S_{Mp} = \Sigma_{a=0}^{p}S_{M0}=-\Sigma_{a=0}^{p}T_{M0}
\int dt Tr(
\Sigma_{m=0}^{9}
\langle[X^{a},X^{b},X^{c}],[X^{a},X^{b},X^{c}]\rangle) =  \nonumber \\ &&
-T_{Mp} \int d^{p+1}\sigma Tr
(\Sigma_{a,b,c=0}^{p}
\Sigma_{i,j,k=p+1}^{10}
\{\langle\partial_{a}X^{i},\partial_{a}X^{i}\rangle
-\frac{1}{4}\langle[X^{i},X^{j},X^{k}],[X^{i},X^{j},X^{k}]\rangle+\nonumber \\ &&\frac{1}{6}
\langle F_{abc},F_{abc}\rangle
\})
\label{a33}
\end{eqnarray}

This is consistent with the earlier studies done on the
Mp-branes  . This mechanism can be applied
for deriving action for other Mp-branes, and so it can be used
for obtaining  suitable action for Dp-branes by  compactifying M-branes on circle. Now, we can answer to question that what is the origin of M0-branes.
 In previous section, we show that one fundamental string decays to D0-anti-D0-branes. On the hand, we obtain the relation between D0brane and M0-brane
 in equation (\ref{a30}). Thus, using equations (\ref{aa14}and\ref{a30}) we can write:

\begin{eqnarray}
&& S_{F-string}= S_{D0}+S_{anti-D0}=S_{M0}+S_{anti-M0}+2V(extra,1)
\label{a34}
\end{eqnarray}

This equation indicates that Fundamental strings can decay to a pair of M0-anti-M0-brane and some extra energy is produced. In fact, the origin of all objects are fundamental strings.

Similar to previous section, we can propose a new mechanism for interaction of two brane-anri-brane. We can write:

\begin{eqnarray}
&& S_{Mp} + S_{anti-Mp}= \Sigma_{a=0}^{p}S_{M0} +
\Sigma_{a=0}^{p}S_{M0}= \nonumber \\&&\Sigma_{a=0}^{p-2}S_{M0} +
\Sigma_{a=0}^{p-2}S_{M0}+ \Sigma_{a=0}^{2}S_{M0}+ S_{M0}=\nonumber
\\&&S_{M(p-2)} + S_{anti-D(M-2)}+S_{M2}+ S_{M0}\label{a35}
\end{eqnarray}

This equation shows that a system of Mp-anti-Mp-brane is broken and two lower dimensional branes and one M2-brane are formed. For example, two M5-branes are broken and two M3-branes and a M2 are produced.

Similar to previous section, the general form of  Born-Infeld
action for Mp-brane  \cite{q17,q19,q21,q22} can be obtain by
replacing two-form commutative brackets with three one  :

\begin{eqnarray}
&& S = - T_{Mp}\int  d^{p+1}\sigma ~ STr \Bigg(-det(P_{abc}[\langle E_{mnl}  ,E_{mik}\rangle(Q^{-1}+\delta)^{ijk}E_{kln}]+\nonumber \\&&  \lambda F_{abc})det(Q^{i}_{j,k})\Bigg)^{1/2}~~
\label{a36}
\end{eqnarray}
where
\begin{eqnarray}
   E_{mnl}^{\alpha,\beta,\gamma} = G_{mnl}^{\alpha,\beta,\gamma} + B_{mnl}^{\alpha,\beta,\gamma}, \qquad  Q^{i}_{j,k} = \delta^{i}_{j,k} + i\lambda[X^{j}_{\alpha}T^{\alpha},X^{k}_{\beta}T^{\beta},X^{k'}_{\gamma}T^{\gamma}]E_{k'jl}^{\alpha,\beta,\gamma} \label{a37}
\end{eqnarray}
 $\lambda=2\pi l_{s}^{2}$, $G_{mnl}=\eta_{mnl}+\partial_{m}X^{i}\partial_{n'}X^{i}\delta^{n'}_{n,l}$ and $X^{i}$ are scalar fields of mass dimension.

Placing two M3-branes at points $u_{1} = l_{2}/2$ and $u_{2} = -l_{2}/2$ and
 one M2-brane between them, we can construct M3-anti-M3-M2-system. Here, $l_{2}$ is the length of M2-brane in transverse direction respect to M3-brane.
  Generalizing  rules in (\ref{a17}) to M-theory, we can achieve to the solutions for  gauge fields in M3 and M2-branes :

\begin{eqnarray}
&&  \lambda F_{011} \text{ in M2-brane } \rightarrow \partial_{t} X^{1} \text{ in M3-brane} \qquad \nonumber \\&& F_{abc} \text{ in M3-brane } \rightarrow [X^{a},X^{b},X^{c}]\text{ in M2-brane }\Rightarrow \nonumber \\&&
X^{i} \sim \frac{l_{3}}{2},\quad A^{i} \sim \frac{l_{2}}{2} \qquad \text{ in M2-brane }
\nonumber \\&& X^{i} \sim \frac{l_{2}}{2},\quad
A^{i} \sim  \frac{l_{3}}{2} \qquad \text{ in M3-brane }
\label{a38}
\end{eqnarray}

where $l_{2}$ and $l_{3}$ are coordinates of M2 and M3-branes respectively.
  Using this equation and equation(\ref{a42}), we can estimate the action for the  case of a  M3-anti-M3-brane pair with lengths $l_{3}$ with a
M2-brane with length $l_{2}$  between them,

\begin{eqnarray}
&& S_{M3-M2-anti-M3}=S_{M3}+S_{anti-M3}+S_{M2}+ S_{M0}\simeq
\nonumber \\ && -T_{M3} \int d^{4}\sigma
(l_{2})^{6}(\sqrt{1+\frac{(l'_{2})^{2}}{4}+\frac{(l'_{3})^{2}}{4}+\frac{(l''_{2})^{2}}{6}+\frac{(l''_{3})^{2}}{6}}+\sqrt{1+\frac{(l'_{2})^{2}}{4}+\frac{(l'_{3})^{2}}{4}+\frac{(l''_{2})^{2}}{6}+\frac{(l''_{3})^{2}}{6}})\nonumber
\\ && -T_{M2} \int d^{3}\sigma
(l_{3})^{6}(\sqrt{1+\frac{(l'_{2})^{2}}{4}+\frac{(l'_{3})^{2}}{4}+\frac{(l''_{2})^{2}}{6}+\frac{(l''_{3})^{2}}{6}})\nonumber
\\ && -T_{M0} \int d\sigma
((l_{2})^{6}+(l_{3})^{6})\sqrt{1+\frac{(l'_{2})^{2}}{4}+\frac{(l'_{3})^{2}}{4}+\frac{(l''_{2})^{2}}{6}+\frac{(l''_{3})^{2}}{6}}\simeq\nonumber
\\ &&-\int dt \frac{(l_{3})^{5}(l_{2})^{4}}{\pi g_{s}l_{s}^{6}}
(\frac{2(l_{2})^{4}}{\pi
l_{s}^{2}}+l_{3}^{3}+(l_{2})^{-4}(l_{3})+(l_{3})^{-5}(l_{2})^{2})(\sqrt{1+\frac{(l'_{2})^{2}}{4}+\frac{(l'_{3})^{2}}{4}+\frac{(l''_{2})^{2}}{6}+\frac{(l''_{3})^{2}}{6}})\simeq\nonumber
\\ && -\int dt
\tilde{F}_{l_{2},l_{3}}(\sqrt{\tilde{D}_{l_{2},l_{3}}})
\label{a40}
\end{eqnarray}

where
\begin{eqnarray}
&& \tilde{F}_{l_{2},l_{3}}=  \frac{(l_{3})^{5}(l_{2})^{4}}{\pi
g_{s}l_{s}^{6}} (\frac{2(l_{2})^{4}}{\pi
l_{s}^{2}}+l_{3}^{3}+(l_{2})^{-4}(l_{3})+(l_{3})^{-5}(l_{2})^{2}),
\nonumber\\ &&
\tilde{D}_{l_{2},l_{3}}=1+\frac{(l'_{2})^{2}}{4}+\frac{(l'_{3})^{2}}{4}+\frac{(l''_{2})^{2}}{6}+\frac{(l''_{3})^{2}}{6}\label{a41}
\end{eqnarray}

and prime($'$) denotes derivative respect to time (t). Now, we can extract the equation of motion from
action (\ref{a40}):
\begin{eqnarray}
&&(\frac{l_{2}'(t)}{\sqrt{1+\frac{(l'_{2})^{2}}{4}+\frac{(l'_{3})^{2}}{4}+\frac{(l''_{2})^{2}}{6}+\frac{(l''_{3})^{2}}{6}}}\acute{)}=\nonumber \\ &&\frac{1}{\sqrt{1+\frac{(l'_{2})^{2}}{4}+\frac{(l'_{3})^{2}}{4}+\frac{(l''_{2})^{2}}{6}+\frac{(l''_{3})^{2}}{6}}}
[\frac{\tilde{F}'_{l_{2},l_{3}}}{\tilde{F}_{l_{2},l_{3}}}(\tilde{D}_{l_{2},l_{3}}-(l_{2}'(t))^{2})]
\label{a42}
\end{eqnarray}
\begin{eqnarray}
&&(\frac{l_{3}'(t)}{\sqrt{1+\frac{(l'_{2})^{2}}{4}+\frac{(l'_{3})^{2}}{4}+\frac{(l''_{2})^{2}}{6}+\frac{(l''_{3})^{2}}{6}}}\acute{)}=\nonumber \\ &&\frac{1}{\sqrt{1+\frac{(l'_{2})^{2}}{4}+\frac{(l'_{3})^{2}}{4}+\frac{(l''_{2})^{2}}{6}+\frac{(l''_{3})^{2}}{6}}}
[\frac{\tilde{F}'_{l_{2},l_{3}}}{\tilde{F}_{l_{2},l_{3}}}(\tilde{D}_{l_{2},l_{3}}-(l_{3}'(t))^{2})]
\label{a43}
\end{eqnarray}

The solutions of these equations are:
\begin{eqnarray}
&& l_{3}\simeq t^{4}\exp(-\frac{(1-e^{\frac{(t-t_{end})^{2}}{\pi^{2} l_{s}^{2}}})}{(t-t_{end})})\nonumber \\ &&l_{2}\simeq t^{5}\exp(-\frac{(1+e^{\frac{(t-t_{end})^{2}}{\pi^{2} l_{s}^{2}}})}{(t-t_{end})})\label{a44}
\end{eqnarray}

where  $t_{end}$ is the age of universe-brane at the end of the present epoch.
 Comparing these solutions with results of equation  (\ref{a22}),
  we find that the value of $l_{2}$ in M-Theory decreases faster than than relevant value in string theory and  M2-brane
   is less stable than D2. The energy momentum tensor for this system is:

 \begin{eqnarray}
&& T^{00}=\tilde{F}_{l_{2},l_{3}}(\sqrt{\tilde{D}_{l_{2},l_{3}}}),  \nonumber \\&&
T^{44}=-\tilde{F}_{l_{2},l_{3}}\frac{1}{\sqrt{\tilde{D}_{l_{2},l_{3}}}}
((l_{3})^{2}(l_{2})^{2}+\frac{(l'_{2})^{2}}{4}) \nonumber \\&& T^{ii} =
-\tilde{F}_{l_{2},l_{3}}\frac{1+(l_{3})^{2}}{3\sqrt{\tilde{D}_{l_{2},l_{3}}}},\,i=1,2,3 \label{a45}
\end{eqnarray}

 Obviously, the energy-momentum tensors in M-Theory increase faster than those in string theory.
  This is because due to this fact that M2-brane dissolves quickly in M3-branes and causes their expansion.
   Similar to string theory, we can write the conservation law in M-Theory.

\begin{eqnarray}
&&\rho
=\rho_{Uni1}+\rho_{Uni2}+\rho_{wormhole}=\rho_{M2-M3}\nonumber
\\&&
p_{i,tot}=p_{i,Uni1}+p_{i,Uni2}+p_{i,wormhole}=p_{i,M2-M3},\text{
   i=1,2,3} \nonumber
\\&& p_{u,tot}=p_{u,wormhole}=p_{z,M2-M3} \label{a46}
\end{eqnarray}
Substituting equations (\ref{a5}), (\ref{a7}) and using (\ref{a45}) and relation (\ref{a24}) in (\ref{a46}), we
derive the following relations:
\begin{eqnarray}
&&6H_{present}^{2}- (\frac{D'}{D}+\frac{C'}{C})\sqrt{\Delta}=
\tilde{F}_{l_{2},l_{3}}\sqrt{\tilde{D}_{l_{2},l_{3}}} \nonumber
\\&&\nonumber
\\&&
2H_{present}^{2} +
4\frac{\ddot{a}_{present}}{a_{present}}+\frac{1}{\sqrt{\Delta}}[2\ddot{V}+2\frac{B'}{B}\dot{V}^{2}+\frac{B'}{B^{2}}+\frac{D'}{D}\Delta]
=\nonumber
\\&&
-\tilde{F}_{l_{2},l_{3}}\frac{1+(l_{3})^{2}}{3\sqrt{\tilde{D}_{l_{2},l_{3}}}}
\nonumber
\\&& \nonumber
\\&&\frac{1}{\sqrt{\Delta}}[2\ddot{V}+2\frac{B'}{B}\dot{V}^{2}+\frac{B'}{B^{2}}+\frac{C'}{C}\Delta]=\nonumber
\\&&-\tilde{F}_{l_{2},l_{3}}\frac{1}{\sqrt{\tilde{D}_{l_{2},l_{3}}}}
((l_{3})^{2}(l_{2})^{2}+\frac{(l'_{2})^{2}}{4}) \label{a47}
\end{eqnarray}

Assuming $D=C$ and $V=l_{2}$ , we can derive the solutions of
these equations:

\begin{eqnarray}
&& a_{present}(t)= e^{-\int dt \ln(W^{-1})},\qquad C(t)=D=
e^{-\int dt \ln(E^{-1})}, \qquad B=e^{-\int dt \ln(R^{-1})}
\nonumber
\\&& \nonumber
\\&& W= [t^{13}\exp(-\frac{(1-e^{\frac{(t-t_{end})^{2}}{\pi^{2} l_{s}^{2}}})}{(t-t_{end})})\exp(-\frac{(1+e^{\frac{(t-t_{end})^{2}}{\pi^{2} l_{s}^{2}}})}{(t-t_{end})})]
[[t^{6}\exp(-\frac{(1+e^{\frac{(t-t_{end})^{2}}{\pi^{2}
l_{s}^{2}}})}{(t-t_{end})})+\nonumber
\\&&t^{7}\exp(-\frac{(1-e^{\frac{(t-t_{end})^{2}}{\pi^{2}
l_{s}^{2}}})}{(t-t_{end})})][1+(4t^{3}\exp(-\frac{(1-e^{\frac{(t-t_{end})^{2}}{\pi^{2}
l_{s}^{2}}})}{(t-t_{end})})+t^{4}(\frac{1}{(t-t_{end})^{2}}+e^{\frac{(t-t_{end})^{2}}{\pi^{2}
l_{s}^{2}}})\times\nonumber
\\&&\exp(-\frac{(1-e^{\frac{(t-t_{end})^{2}}{\pi^{2}
l_{s}^{2}}})}{(t-t_{end})}))^{2}+
5t^{4}\exp(-\frac{(1+e^{\frac{(t-t_{end})^{2}}{\pi^{2}
l_{s}^{2}}})}{(t-t_{end})})+
t^{5}(\frac{1}{(t-t_{end})^{2}}-e^{\frac{(t-t_{end})^{2}}{\pi^{2}
l_{s}^{2}}})\times\nonumber
\\&&\exp(-\frac{(1+e^{\frac{(t-t_{end})^{2}}{\pi^{2}
l_{s}^{2}}})}{(t-t_{end})}))^{2}]^{1/4}\nonumber
\\&& \nonumber
\\&& E=[t^{6}\exp(-\frac{(1+e^{\frac{(t-t_{end})^{2}}
{\pi^{2}
l_{s}^{2}}})}{(t-t_{end})})+t^{7}(\frac{1}{(t-t_{end})^{2}}-
e^{\frac{(t-t_{end})^{2}}{\pi^{2}
l_{s}^{2}}})\exp(-\frac{(1+e^{\frac{(t-t_{end})^{2}}{\pi^{2}
l_{s}^{2}}})}{(t-t_{end})})] \times\nonumber \\
&&[t^{13}\exp(-\frac{(1-e^{\frac{(t-t_{end})^{2}}{\pi^{2}
l_{s}^{2}}})}{(t-t_{end})})\exp(-\frac{(1+e^{\frac{(t-t_{end})^{2}}{\pi^{2}
l_{s}^{2}}})}{(t-t_{end})})]
[[t^{6}\exp(-\frac{(1+e^{\frac{(t-t_{end})^{2}}{\pi^{2}
l_{s}^{2}}})}{(t-t_{end})})+\nonumber
\\&&t^{7}\exp(-\frac{(1-e^{\frac{(t-t_{end})^{2}}{\pi^{2}
l_{s}^{2}}})}{(t-t_{end})})][1+(4t^{3}\exp(-\frac{(1-e^{\frac{(t-t_{end})^{2}}{\pi^{2}
l_{s}^{2}}})}{(t-t_{end})})+t^{4}(\frac{1}{(t-t_{end})^{2}}+e^{\frac{(t-t_{end})^{2}}{\pi^{2}
l_{s}^{2}}})\times\nonumber
\\&&\exp(-\frac{(1-e^{\frac{(t-t_{end})^{2}}{\pi^{2}
l_{s}^{2}}})}{(t-t_{end})}))^{2}+
5t^{4}\exp(-\frac{(1+e^{\frac{(t-t_{end})^{2}}{\pi^{2}
l_{s}^{2}}})}{(t-t_{end})})+
t^{5}(\frac{1}{(t-t_{end})^{2}}-e^{\frac{(t-t_{end})^{2}}{\pi^{2}
l_{s}^{2}}})\times\nonumber
\\&&\exp(-\frac{(1+e^{\frac{(t-t_{end})^{2}}{\pi^{2}
l_{s}^{2}}})}{(t-t_{end})}))^{2}]^{-1/2}\nonumber
\\&& \nonumber
\\&& R=[1+t^{6}\exp(-\frac{2(1-e^{\frac{(t-t_{end})^{2}}{\pi^{2} l_{s}^{2}}})}{(t-t_{end})})]\times \nonumber\\&&[t^{13}\exp(-\frac{(1-e^{\frac{(t-t_{end})^{2}}{\pi^{2} l_{s}^{2}}})}{(t-t_{end})})\exp(-\frac{(1+e^{\frac{(t-t_{end})^{2}}{\pi^{2} l_{s}^{2}}})}{(t-t_{end})})]
[[t^{6}\exp(-\frac{(1+e^{\frac{(t-t_{end})^{2}}{\pi^{2}
l_{s}^{2}}})}{(t-t_{end})})+\nonumber
\\&&t^{7}\exp(-\frac{(1-e^{\frac{(t-t_{end})^{2}}{\pi^{2}
l_{s}^{2}}})}{(t-t_{end})})][1+(4t^{3}\exp(-\frac{(1-e^{\frac{(t-t_{end})^{2}}{\pi^{2}
l_{s}^{2}}})}{(t-t_{end})})+t^{4}(\frac{1}{(t-t_{end})^{2}}+e^{\frac{(t-t_{end})^{2}}{\pi^{2}
l_{s}^{2}}})\times\nonumber
\\&&\exp(-\frac{(1-e^{\frac{(t-t_{end})^{2}}{\pi^{2}
l_{s}^{2}}})}{(t-t_{end})}))^{2}+
5t^{4}\exp(-\frac{(1+e^{\frac{(t-t_{end})^{2}}{\pi^{2}
l_{s}^{2}}})}{(t-t_{end})})+
t^{5}(\frac{1}{(t-t_{end})^{2}}-e^{\frac{(t-t_{end})^{2}}{\pi^{2}
l_{s}^{2}}})\times\nonumber
\\&&\exp(-\frac{(1+e^{\frac{(t-t_{end})^{2}}{\pi^{2}
l_{s}^{2}}})}{(t-t_{end})}))^{2}]^{-1/2}\label{a48}
\end{eqnarray}

 As can be seen from this equation, the parameters of wormhole and
 scale factor of universe are zero at t=0, increase with time,
 turn over a maximum and reduce to zero at the end. From this
 point of view, we will have a birth time for universe. However,
 regarding initial transitions, we can show that universe has no
 beginning. Similar to string theory, we define a new scale factor
$a_{tot}=a_{initial}+a_{present}$ which $a_{initial}$ can be
obtained from following equation:

\begin{eqnarray}
&&\rho_{\text{fundamental string}}=\rho_{3M0}+ \rho_{3-anti-M0}=
\rho_{uni1}+ \rho_{uni2}\Rightarrow \nonumber
\\&&6H_{initial}^{2} = 2T_{M0}
\int dt Tr( \Sigma_{M,N,L=0}^{10}
\langle[X^{M},X^{N},X^{L}],[X^{M},X^{N},X^{L}]\simeq \nonumber
\\&&\frac{t^{9}}{9}\exp(-\frac{2(1-e^{\frac{(t-t_{end})^{2}}{\pi^{2} l_{s}^{2}}})}{(t-t_{end})})+ \frac{(t-t_{end})^{8}}{8}
+\frac{(t-t_{end})^{2}}{2}e^{\frac{(t-t_{end})^{2}}{\pi^{2}
l_{s}^{2}}}\rightarrow \nonumber
\\&& a_{initial}(t)=Aexp(-\int_{t_{f-string}}^{t} dt \frac{t^{9}}{9}\exp(-\frac{2(1-e^{\frac{(t-t_{end})^{2}}{\pi^{2} l_{s}^{2}}})}{(t-t_{end})})+ \frac{(t-t_{end})^{8}}{8}
+\frac{(t-t_{end})^{2}}{2}e^{\frac{(t-t_{end})^{2}}{\pi^{2}
l_{s}^{2}}})
 \label{aa25}
\end{eqnarray}

Here, $t_{f-string}$ is the age of the fundamental string and
$X^{2}=l_{2}$ and $X^{3}=l_{3}$.  An interesting result that comes
out of this equation is that only in the case of
($t_{f-string}\rightarrow \infty$), the initial scale factor is
zero. This means that the age of fundamental string is infinity
and since this string is the initial state of present universe, we
conclude that the total age of universe is infinity.

\section{Summary and Discussion} \label{sum}
In this paper, we have reconsidered the results of \cite{q1} in
string theory and M-theory. We have shown that there is no big
bang for our universe in string theory and the age of universe is
infinity which is in agreement with predictions of \cite{q1}. We
have discussed that universe was a fundamental string at the
beginning. this string decayed to N D0 and anti-D0-brane. These
branes joined to each other and formed a system of
D5-anti-D5-brane. The brane and antibrane interact with each other
with exchanging D0-branes and transit to lower dimensional D3 and
anti-D3-brane. Our universe is located on one of D3-branes and
emitted D0-branes constructed a wormhole. This wormhole connects
two universes and causes to evolution of scale factor of present
form of  universe. We have observed that this scale factor is zero
at the birth of D3-branes. Then, we define a new scale factor in
related to initial states of universe. This scale factor is zero
only in the case that only in the case of
($t_{f-string}\rightarrow \infty$) where $t_{f-string}$ is the age
of fundamental string. This means that the age of fundamental
string is infinity and since this string is the initial state of
present universe, we conclude that the total age of universe is
infinity.

\section*{Acknowledgments}
A.Sepehri would like to thank of Shahid Bahonar University of
Kerman for financial support during investigation in this work.
Also, he wishes to thank of Ali Mohammad for his lecture in
cosmology that gives new insight in this subject.

 \end{document}